\documentstyle[prb,aps,multicol,epsfig]{revtex}

\begin{document} \bibliographystyle{unsrt} 
\input{epsf}
\title{Dynamical studies of the response function in a Spin Glass}
\author
{
K. Jonason and P. Nordblad
}
\address
{
Dept. of Materials Science, Uppsala University, Box 534, S-751 21 Uppsala,
Sweden
}

\maketitle

\begin{abstract}

Experiments on the time dependence of the response function of a
Ag(11 at$\%$Mn) spin glass at a temperature below the zero field spin
glass temperature are used to explore the non-equilibrium nature of
the spin glass phase.  It is found that the response function is only
governed by the thermal history in the very neighbourhood of the
actual measurement temperature.  The thermal history outside this
narrow region is irrelevant to the measured response.  A result that
implies that the thermal history during cooling (cooling rate, wait
times etc.)  is imprinted in the spin structure and is always
retained when {\it any} higher temperature is recovered.  The observations are
discussed in the light of a real space droplet/domain phenomenology.
The results also emphasise the importance of using controlled cooling
procedures to acquire interpretable and reproducible experimental
results on the non-equilibrium dynamics in spin glasses.

\end{abstract}

\bigskip
\begin{multicols}{2}
\narrowtext

\section{introduction}

The non-equilibrium nature of the spin glass phase has been
extensively investigated since it was discovered in low frequency
ac-susceptibility measurements \cite{Lgrac} and dc-relaxation
experiments in the early 1980's \cite{Lgr}.  The interpretation of the
experimental results and the design of new experimental procedures
emanate essentially from two different theoretical approaches:
hierarchical phase space models \cite{Saclay} and real space
droplet/domain models \cite{three,henke}.  A high level of
phenomenlogical and theoretical insight into the phenomena has by now
been acquired.  There remain, however, unresolved problems as to the
interpretation and reproducibility of non-equilibrium results.  These
shortcomings do in part also prevent a useful judgement in-between the
applicability of the different models to real 3d spin glasses.  The
recently reported memory effect\cite{one,two}, observable in low
frequency ac-susceptibility experiments, not only elucidates some
paradoxal features of the spin glass phase, but an in detail study of
related phenomena \cite{four} also emphasizes the importance of:
cooling/heating rates, wait times, thermalisation times etc., i.e the
detailed thermal history of the sample on the results from low
frequency ac-susceptibility and dc-magnetic relaxation experiments.
These complicated non-equilibrium phenomena should be considered in
the perspective that an ac-experiment at only a couple of decades
higher frequency, $f\geq$ 100 Hz (observation time 1/$\omega\leq$ 2
ms), in spite of a strong frequency dependence, simply shows an
equilibrium character when measured in ordinary ac-susceptometers.
The non-equilibrium processes of the system are only active on
observation time scales governed by the cooling/heating rate and
during a halt at constant temperature these evolve with the time the
sample has been kept at constant temperature.

In this paper we report results from low field zero field cooled (zfc)
relaxation experiments, i.e.  measurements of the response function,
at one specific temperature in the spin glass phase.  The parameter we
vary in a controlled

\begin{figure}

\centerline{\hbox{\epsfig{figure=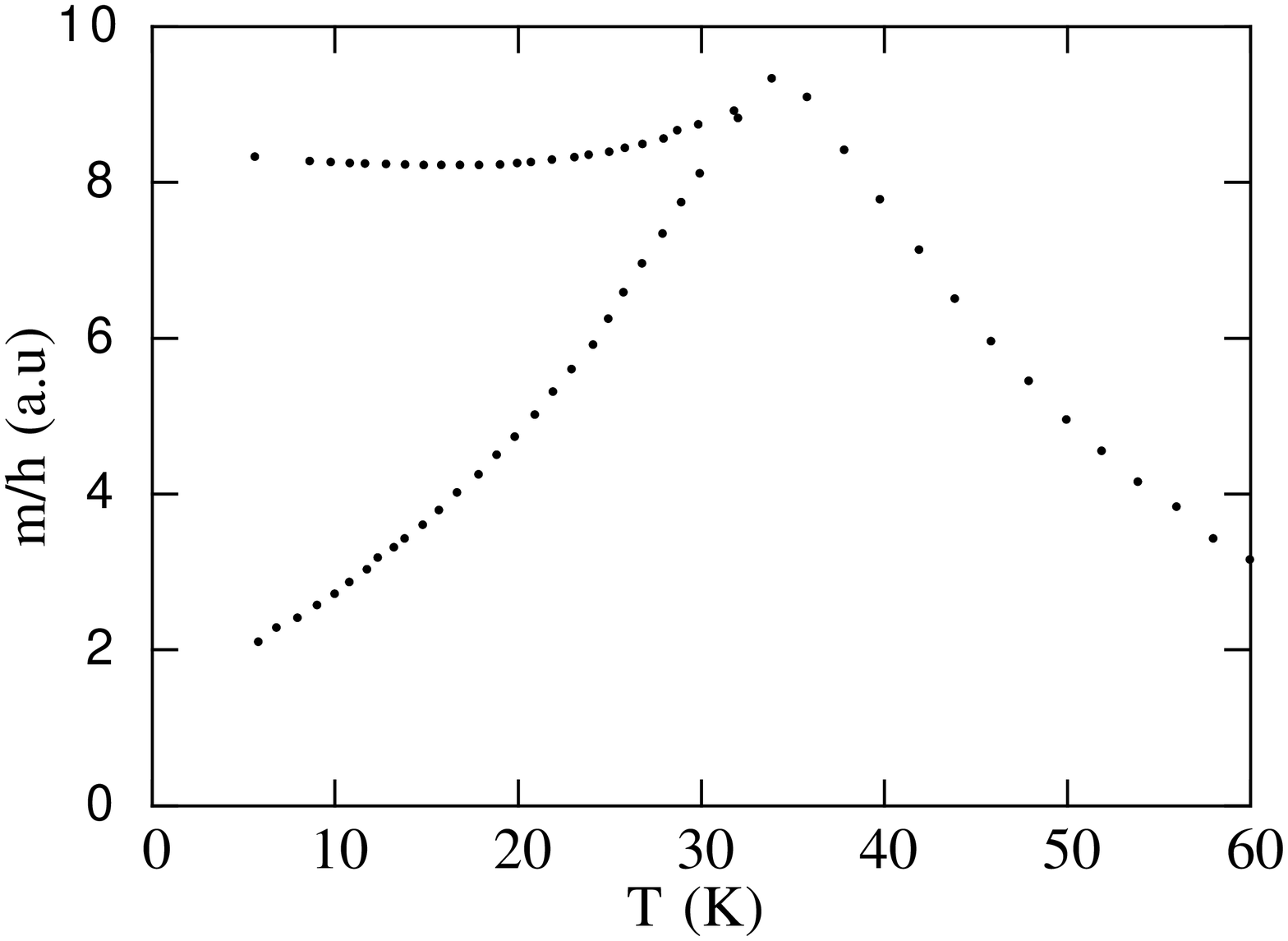,width=8.5cm}}}
\caption{
\hbox {Fc- and zfc-magnetisation of Ag(11 at$\%$Mn) }plotted vs.
temperature in an applied field of 40 Oe.
}
\label{fig1}
\end{figure}

 way is the thermal history in the spin glass
phase.  The results imply that the reponse function is governed by:
the cooling procedure in a rather narrow region just above the
measurement temperature, the wait time before the response function is
measured, and the cooling/heating procedure in a rather narrow region
just below the measurement temperature if an additional undercooling
has been carried out.  It is also clearly shown that the thermal history,
in the spin glass phase, at temperatures well enough separated from
the measurement temperature is irrelevant to the reponse function at
$T_{m}$.  The results are put in relation to a phenomenological real
space fractal domain picture that is introduced and discussed in more
detail elsewhere \cite{four,interference}.

The investigation is also motivated by a lack of agreement in the
detailed behaviour of the non-equilibrium dynamics, when measured on
one and the same spin glass material in different magnetometers.

\begin{figure}

\centerline{\hbox{\epsfig{figure=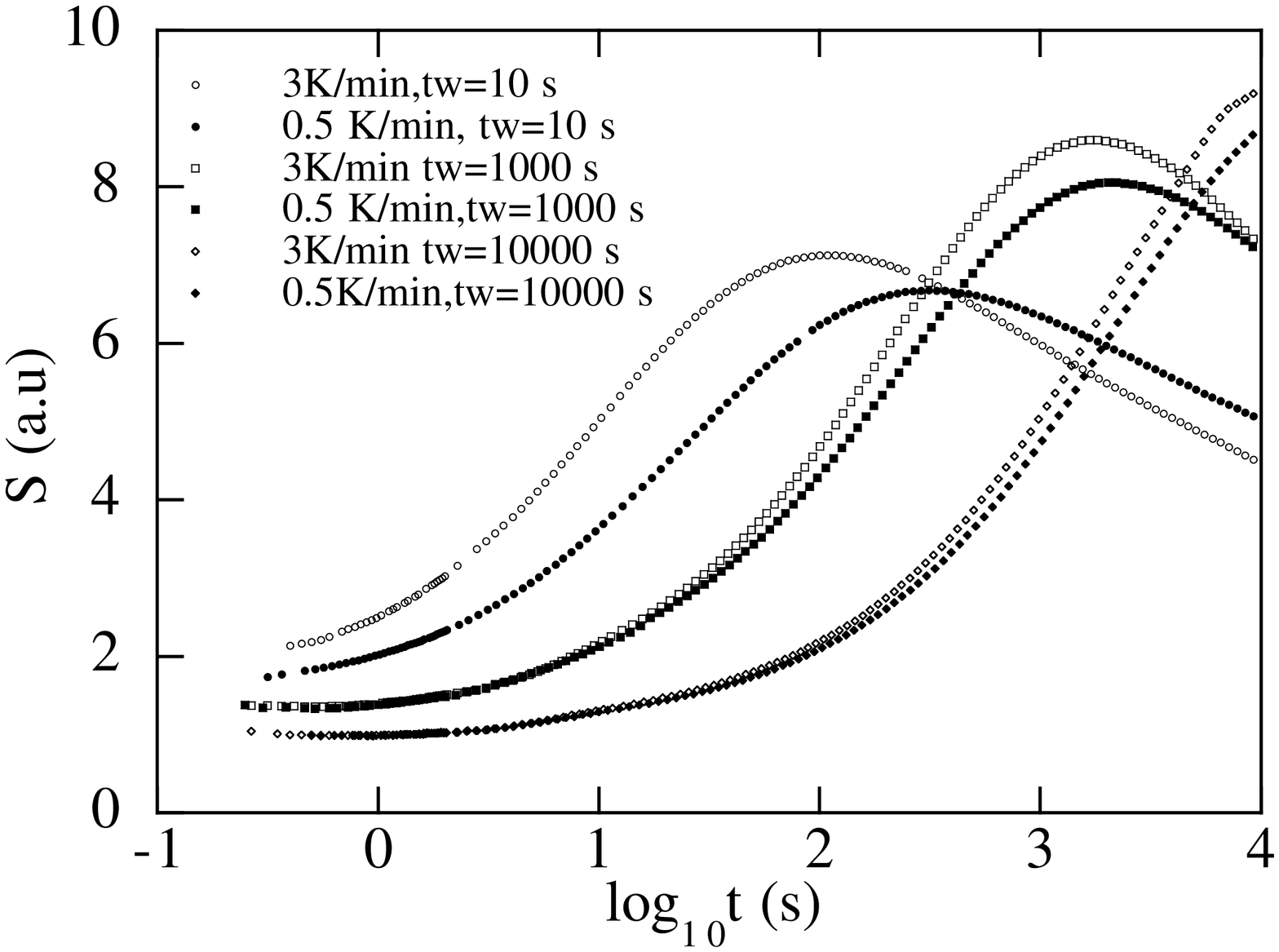,width=8.5cm}}}
\caption{
\hbox {The relaxation rate, $S(t)=1/h$ $\partial
m$$/$$\partial$ln$t$} vs.  time, measured at $T_{m}$ = 27 K for three
different wait times, $t_{w}$, as indicated in the figure, using two
different cooling rates 3K/min.  and 0.5 K/min, $h$=1 Oe.
}
\label{fig2}
\end{figure}

\section{experimental}

The sample is a bulk piece of Ag(Mn 11at$\%$) with a spin glass
temperature $T_{g}\approx$ 35 K made by drop-synthesis technique.  The
experiments were performed in a non-commercial SQUID magnetometer
optimised for dynamic studies in low magnetic fields \cite{five}.  The
sample is cooled in zero magnetic field from a temperature above
$T_{g}$, using a controlled thermal sequence, to the measurement
temperature, $T_{m}$= 27 K, where a weak magnetic field ($h$= 1 Oe) is
applied after the system been kept at $T_{m}$ a certain wait time,
$t_{w}$.  The relaxation of the magnetisation is then recorded as a
function of time elapsed after the field application.  In our figures
of the relaxation data we show the relaxation rate, i.e.  the
logarithmic derivative of the response function $S$=1/$h$
d$m$/dlog$t$, which is the quantity that most clearly exposes changes
of the response function after the different thermal procedures.  The
relaxation rate is related to the out-of-phase component of the
ac-susceptibility via $S$($t$)$\approx$-2/$\pi$ $\chi''$($\omega$) at
$t$=1/$\omega$ \cite{six}, and $\chi''(T)$ is also the quantity that
most instructively has been used to visualise the memory phenomenon in
spin glasses mentioned above \cite{two,four}.

To acquaint with the sample, the field cooled and zero field cooled
magnetisation is plotted vs.  temperature in Fig.  1.  The curves are
measured in a field of 40 Oe.  The cusp in the zfc susceptibility
around 35 K closely reflects the spin glass temperature $T_{g}$ of the
sample.

\section{Results}

The classic aging experiment is performed by cooling the sample
directly to a measurement temperature below $T_{g}$, wait a controlled
amount of time and then switch the magnetic field on or off.  Not much
attention has been given to the influence of the cooling rate.  In
Fig.  2, the relaxation rate, $S(t)$, is plotted vs.  log$t$ for three
different wait times.  The results for two substantially different,
but still in a logarithmic time perspective rather similar, cooling
rates are plotted, 3 K/min open symbols and 0.5 K/min solid symbols.
The wait time dependence of the data displays the characteristics of
the ageing phenomenon in spin glasses (a similar ageing phenomenon is
also an inherent property of other disordered and frustrated magnetic
systems \cite{seven}).  The response is visually sensitive to the
cooling rate, $t_{c}$, but the influence of $t_{c}$ decreases as the
wait time increases.  For the $t_{w}$=10 s curves, rapid cooling
yields a maximum in $S$($t$) at a shorter observation time than slow
cooling.  The two curves are clearly different and do not coalesce
anywhere in the measured time interval.  For the longer wait times,
the position of the maxima do not differ appreciably, but the
magnitude is somewhat higher for the rapidly cooled curves.  At short
observation times, for the $t_{w}$=$10^3$ and $10^4$ s curves, there
are no or weak differences in the relaxation rates, but after some
time they start to deviate and there remains a cooling rate dependence
all through the long time part of our observation time window.

Concentrating on the position of the maximum in the relaxation rate
and identifying this with an effective age of the system, $t_{aeff}$,
this parameter is apparently governed by the cooling rate and the wait
time.  When the wait time is short, $t_{aeff}$ is governed by the
cooling rate, whereas for longer wait times, it is dominated by the
wait time, $t_{w}$.  The position of the maximum in the relaxation
rate is closely equal to the wait time for $t_{w}$ = $10^3$ and $10^4$
s while for the $t_{w}$ = 10 s the maximum is delayed about a decade
in time.  Similar tendencies as to the evolution of the effective age
with increasing wait time have earlier been reported in connection
with a specific method, `field quenching', to achieve a well defined
initial state for ageing experiments \cite{fieldquenching}.

To further investigate how the thermal sequence on approaching the
measurement temperature influences the response function, temperature
shift experiments under controlled cooling and heating rates were
performed.  In the negative temperature shift experiment, the system
is kept a certain wait time at a shift temperature, $T_{s}$, above
$T_{m}$.  Therafter the system is cooled to the measurement
temperature where the field is applied immediatly after the sample has
become thermally stabilised.  The positive temperature shift
experiment follows a similar procedure, but the temperature $T_{s}$ is
below $T_{m}$.  Fig.  3 shows the relaxation rate when the system has
been subjected to (a) negative and (b) positive temperature shifts.
The wait time at $T_{s}$ is 1000 s and $T_{m}$ = 27 K. The cooling and
heating rates are 0.5 K/min.  Two reference curves measured after the
sample has been cooled at 0.5 K/min directly to $T_{m}$ are included
in the figure.  One corresponds to a wait time $t_{w}$ = 1000 s and
the other is measured without any wait time prior to the field
application ($t_{w}\approx$ 0).  From Fig.  3(a) it is seen that for
$T_{s}$ $>$ 30 K the response is indistinguishable from what is
measured if the sample

\begin{figure}

\centerline{\hbox{\epsfig{figure=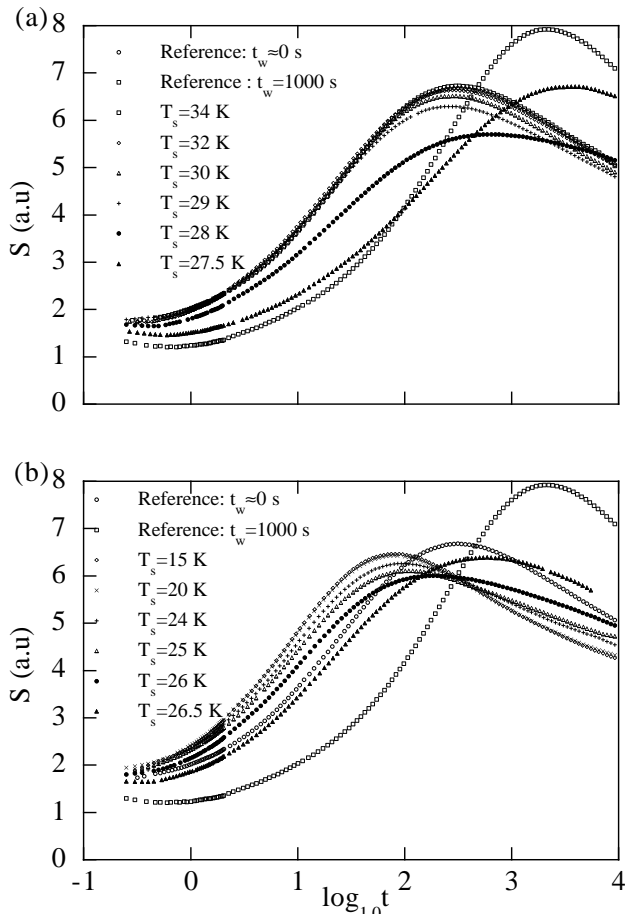,width=8.5cm}}}
\caption{
\hbox {The relaxation rate, $S(t)$ vs.  time, employing a} halt
lasting 1000 s in the cooling procedure at different temperatures (a)
$T_{s}$ = 34, 32, 30, 29, 28 and 27.5 K above $T_{m}$ = 27 K and at
different temperatures (b) $T_{s}$ = 15, 20, 24, 25, 26 and 26.5 K below
$T_{m}$ = 27 K (b).  In both (a) and (b) two reference curves are
included that are measured after immediately cooling the sample to
$T_{m}$ at a cooling rate of 0.5 K/min. and using $t_{w}\approx$ 0  and 1000
s,  $h$= 1 Oe.
}
\label{fig3}
\end{figure}

 is cooled directly to $T_{m}$.  The time the
sample has been kept at $T_{s}$ is irrelevant for the response at
$T_{m}$.  When the shift temperature $T_{s}$ is lower than 30 K, the
time spent at the higher temperature starts to influence the response
at $T_{m}$.  The first deviation from the reference ($t_{w}\approx$ 0)
curve occurs at long observation times, the maximum in the relaxation
rate decreases in magnitude and on further decreasing $T_{s}$ it
shifts towards longer times, and finally when $T_{s}$ approaches
$T_{m}$, the relaxation rate approaches the $t_{w}$ = 1000 s reference
curve.  The important implication of this behaviour is that only the
thermal history in a limited temperature region just above $T_{m}$
governs the reponse function.

The results from positive temperature shifts shown in Fig.  3(b) give
a somewhat more complicated result.  A first observation is that when
the system has been cooled to and aged at temperatures well below
$T_{m}$, the measured curves are identical, but different from the
$t_{w}\approx$ 0 reference curve, the maximum in the relaxation rate

\begin{figure}

\centerline{\hbox{\epsfig{figure=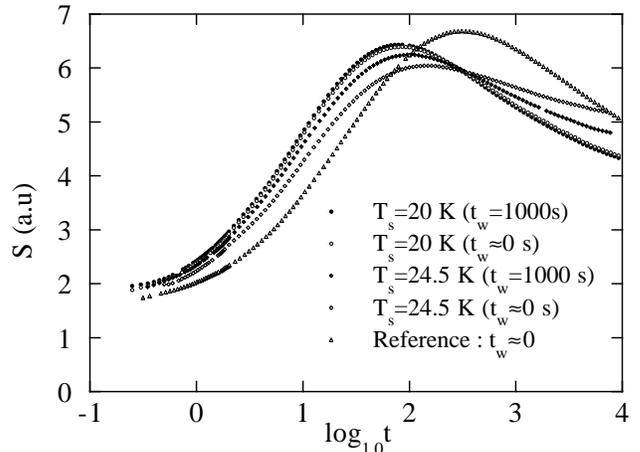,width=8.5cm}}}
\caption{
\hbox {The relaxation rate $S(t)$ vs.  time, after cooling} the
sample to a low temperature $T_{s}$ = 20 and 24.5 K, halt at this
temperature for 0 s or 1000 s and then approach $T_{m}$ = 27 K at a
heating rate of 0.5 K/min.  where the field is immediately applied.  A
reference curve $t_{w}\approx$ 0 measured after cooling the sample
directly to $T_{m}$ without any halt is also plotted, $h$ = 1 Oe
}
\label{fig4}
\end{figure}

occurs
 at a shorter time than for the reference curve.  Thus, the
response function is somewhat different after only cooling to the
measurement temperature at one specific cooling rate, compared to
after substantially undercooling the sample and re-heat it with a
similar heating rate.  The curves representing $T_{s}$= 15 and 20 K are
indistiguishable from each other, but when $T_{s}$ is increased
further, the relaxation rate gets surpressed and the maximum shifts
towards longer times to finally coalesce with the reference $t_{w}$=
1000 s curve when $T_{s}\approx$ $T_{m}$.  Fig.  4 shows relaxation
rate curves from experiments where the sample is undercooled to
$T_{s}$, and either kept there 1000 s (as in Fig.  3(b)) or
immediately re-heated to $T_{m}$, where the relaxation is measured
using $t_{w}\approx$ 0.  The cooling/heating rates are again 0.5
K/min.  The figure shows that the wait time at the lowest temperature
does not affect the results, identical curves are measured whether the
sample is kept at 20 K for 1000 s or 0 s.  For a shift temperature,
$T_{s}$= 24.5 K, closer to $T_{m}$, a clear difference is observed
between the two curves with different wait times.  The implication
from these data is again that the thermal history far enough away from
the measurement temperature is irrelvant to the response function at
$T_{m}$, the response is in the experimental time window fully
governed by the previous cooling/heating history in the very
neighbourhood of $T_{m}$.

In Fig.  5, results for (a) negative and (b) positive temperature
shifts using two different cooling rates 3 K/min and 0.5 K/min (as in
Fig.  3) are displayed.  The cooling/heating rate from $T_{s}$ to
$T_{m}$ = 27 K is always the same, 0.5 K/min.  For negative temperature
shifts (Fig.  5 (a)) there is no influence of the different cooling
rates for $T_{s}$ $>$ 29 K. The rapid and slow cooling give the same
response at the measurement temperature. For $T_{s}$ = 28 K influences
of the different cooling rates start to become
observable at
observation times, $t$ $>10^3$ s.  At shorter observation times no sign
of the different cooling rates can be resolved.  When $T_{s}$ gets
even closer to $T_{m}$, the influence from the differences in cooling
rates appears earlier, but still there is a region at shorter
observation times where the response is independent of the cooling
rate.

For positive temperature shifts, Fig.  5 (b), the result is different.
There are clear cooling rate effects for all different temperatures
$T_{s}$.  The curve representing rapid cooling always has a larger
magnitude and a sharper maximum that occurs at shorter observation
times than the corresponding slowly cooled curve.  This result
implies that the cooling rate in the region just above $T_{m}$ always
remains one of the governing parameters for the response after
undercooling the sample, i.e.  a memory of this cooling process
becomes imprinted in the spin structure and is conserved in spite of
the re-structuring that occurs at lower temperatures.

\section{Discussion}

Our current results on the non-equilibrium response function show that
the cooling rate significantly influences the measured response
function, especially the reponse at short wait times is dominated by
the specific cooling rate employed.  Also, in experiments where the
sample has been undercooled below the measurement temperature, the
cooling rate above the measurement temperature remains one of the
governing parameters for the non-equilibrium response.  These findings
have practical importance for the design of experimental procedures
and when making detailed comparisons between results obtained in
different experimental set-ups and in different laboratories on
similar spin glass materials.

We have, using different experimental procedures, elucidated an
apparently paradoxal property of the non-equilibrium spin structure in
spin glasses: the spin structure records the cooling history, this
thermal history (cooling/heating rate, wait times etc.)  remains
imprinted in the configuration, and the fragment of the thermal
history confined in a rather narrow region close to $\it any$ higher
temperature $T_{m}$ within the spin glass phase, can be recovered in a
relaxation experiment at $T_{m}$.  A continuous memory recording
occurs on cooling, in spite of the fact that the spin structure is
subjected to substantial reconfiguration at all temperatures below
$T_{g}$ as is observed from the ever present ageing behaviour; and
although the short time response appears equilibrated at all
temperatures in the spin glass phase.

Employing the droplet scaling model \cite{three} and the concepts
chaos and overlaplength \cite{eight}, the observations of
non-equilibrium and ageing behaviour observed after cooling the sample
to a temperature in the spin glass phase can be accounted for as an
immediate consequence of the growth of the size of correlated spin
glass domains at 

\begin{figure}

\centerline{\hbox{\epsfig{figure=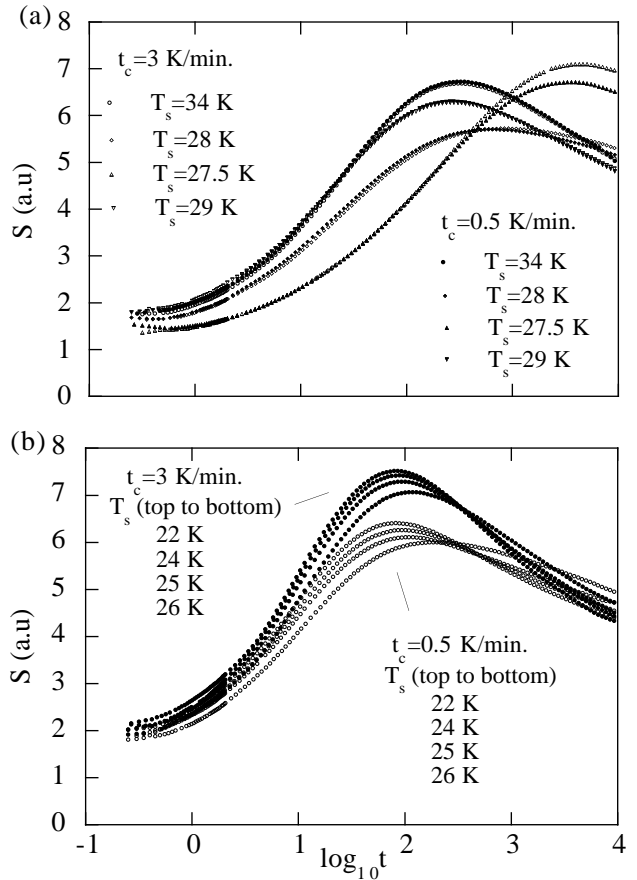,width=8.5cm}}}
\caption{
\hbox {The relaxation rate, $S(t)$ vs.  time, cooling the} sample at
two different rates, 3 K/min.  and 0.5 K/min., employing a halt
lasting 1000 s in the cooling procedure at different high temperatures
(a) $T_{s}$ = 34, 29, 28 and 27.5 K before approaching $T_{m}$ = 27 K
with a cooling rate of 0.5 K/min., and at different low temperatures
(b) $T_{s}$ = 15, 22, 24, 25 and 26 K before approaching $T_{m}$
= 27 K with a heating rate of 0.5 K/min. The magnetic field $h$=1 Oe is
applied immediately when reaching $T_{m}$.
}
\label{fig5}
\end{figure}

constant temperature.  Interpretations of ageing in
these terms have been extensively discussed in numerous articles and
are recently reviewed in ref.\cite{one}.  A somewhat modified version
of the droplet scaling model that has been constructed to account for
the memory behaviour\cite{two} is extensively discussed in ref.
\cite{four}.

The continuous memory recording of the thermal history that is
exemplified by the current experimental results does however require a
comment on the paradoxal possibility that reconfiguration on different
length scales are fully separable, i.e.  that reconfiguration of the
spin structure on short length scales allows a simultaneous stability
of the old configuration on larger length scales.  The droplet model
prescribes one unique equilibrium spin glass state with time reversal
symmetry and that domains of spin glass ordered regions grow
unrestrictedly with time at constant temperature.  If the temperature
is altered, the equilibrium configuration also alters due to chaos,
but there is an overlap on short length scales between the equilibrium
configurations, the length scale of which rapidly decreases with
increasing separation between the two temperatures.  The experimental
results show that a memory of the spin structure that has developed at
high temperatures remains imprinted in the non-equilibrium structure
at lower temperatures (but also that it is rapidly erased if the
temperature is increased above the original aging temperature).  Such
a re-stored spin structure requires that all reconfiguration at lower
temperatures must occur only on small lengths scales and in dispersed
regions.  The bulk of the numerous droplet excitations of different
sizes that do occur may not cause irreversible changes of the spin
structure on large length scales, but are to be excited within already
equilibrated regions of spin glass order.

How can this rather abstract picture of the dynamic spin structure be
related to our current experimental observations?  The processes that
cause the increase of the magnetisation in a zfc magnetic relaxation
experiment is a polarisation of spontaneous droplet excitations.  The
measured quantity, the zfc magnetisation, gives an integrated value of
polarisation of all droplet excitation with relaxation time shorter
than the observation time and the measured relaxation corresponds to
droplets with relaxation time of the order of the observation time,
$t$.  In an ac-susceptibility experiment, the in-phase component gives
the integrated value of all droplet excitations with relaxation time
shorter than the observation time, $t$ = 1/$\omega$, and the out-of
phase component measures the actual number of droplets of relaxation
time equal to 1/$\omega$.  The non-equilibrium characteristics imply
that the distribution of droplet excitations changes with the time
spent at constant temperature and that there is an excess of droplet
excitations of a size that corresponds to a relaxation time of the
order of the wait time.  On shorter time scales an equilibrium
distribution has been attained, reflected by the equilibrium reponse
always obtained in ac-susceptibility experiments at higher
frequencies, and in zfc measurements at long but different wait times
by the fact that a similar (equilibrium) reponse is approached at the
shortest observation times.  The implication of the cooling rate
dependence is that the actual distribution of active droplets is
governed by the cooling rate and the wait time at constant
temperature, and that this distribution in turn is governed by the
underlying spin configuration.  The phenomenon that the sample retains
a distribution of droplets that is governed by the cooling rate and
any previous wait time at the measurement temperature, then implies
that a closely equivalent spin configuration to the original one is
also retained when the temperature is recovered.

\section{conclusions}

We have shown that non-equilibrium dynamics measured at a specific
temperature in spin glasses is primarily governed by the thermal
history close to this temperature during the cooling sequence.  If the
sample has been undercooled, the response is also partly affected by
the heating rate towards the measurement temperature.  The behaviour
may be incorporated in a real space picture of a random spin
configuration containing fractal spin glass domains of sizes that
increase through spontaneous droplet excitations.

The results emphasise the importance of well controlled experimental
procedures when studying non-equilibrium dynamics of spin glasses.

\section{acknowledgments}

Financial support from the Swedish Natural Science Research Council
(NFR) is acknowledged.  Numerous and useful discussions on the memory
phenomenon and the non-equilibrium nature of the spin glass phase with
T. Jonsson, E. Vincent, J.-P. Bouchaud and J. Hammann are acknowledged.

\end{multicols}


\begin {references}

\bibitem{Lgrac} L. Lundgren, P. Svedlindh and O. Beckman, J. Magn.
Magn. Mater. {\bf 31-34}. 1349 (1983).

\bibitem{Lgr} L. Lundgren, P. Svedlindh, P. Nordblad and O. Beckman,
Phys.  Rev.  Lett.  {\bf 51}, 911 (1983).

\bibitem{Saclay} see e.g.  E. Vincent, J.P. Bouchaud, J. Hammann and
F. Lefloch, Phil.  Mag.  B {\bf 71} (1995); J.-P. Bouchaud, L.F.
Cugliandolo, J. Kurchan and M. M\'ezard, in {\it Spin Glasses and
Random Fields} p 161-223, ed.  A. P. Young, World Scientific (1998).

\bibitem{three} D. S. Fisher and D. A. Huse, Phys. Rev. B {\bf 38},
373 (1988); {\bf 38}, 386 (1988).

\bibitem{henke} G.J.M. Koper and H.J. Hilhorst, J. Phys. (France) {\bf
49}, 429 (1988).

\bibitem{one} P. Nordblad and P. Svedlindh; in {\it Spin Glasses and
Random Fields} p 1-27, ed.  A. P. Young, World Scientific (1998).

\bibitem{two} K. Jonason, E. Vincent, J. Hammann, J. P. Bouchaud, P.
Nordblad, Phys. Rev. Lett. {\bf 81}, 3243 (1998).

\bibitem{four} T. Jonsson, K. Jonason, P. J\"{o}nsson and P. Nordblad, Phys.
Rev. B {\bf 55}, 8770 (1999).

\bibitem{interference} K. Jonason, P. Nordblad, E. Vincent, J. Hammann
and J.P. Bouchaud, unpublished.

\bibitem{five} J. Magnusson, C. Djurberg, P. Granberg, P. Nordblad
Rev. Sci. Instrum. {\bf 68}, 3761 (1997).

\bibitem{six} L. Lundgren, P. Svedlindh and O. Beckman, J. Magn. Magn.
Mater. {\bf 25}, 33 (1981).

\bibitem{seven} P. Nordblad; in {\it Dynamical Properties of
Unconventional Magnetic Systems} p 343-366, eds.  A.T. Skjeltorp and
D. Sherrington, Kluwer (1998).

\bibitem{fieldquenching} P. Nordblad, P. Svedlindh, P. Granberg and L.
Lundgren, Phys. Rev. B {\bf 35}, 7150 (1987).

\bibitem{eight}  A. J. Bray and M. A. Moore, Phys. Rev. Lett. {\bf
58}, 57 (1987).



\end{references}
\end{document}